\pdfoutput=1
\documentclass[11pt,twoside,a4paper,cmspaper,final,collab]{cms-tdr}
\def\svnVersion{4241e22}\def\svnDate{2025/07/16}\def\cmsCernNoTag{CERN-EP-2025-102}\def\cmsCernDate{\today}\def\cmsMessage{Published in Physical Review Letters as \href{http://dx.doi.org/10.1103/6z3d-zjw4}{\doi{10.1103/6z3d-zjw4}.}}
\begin{document}\cmsNoteHeader{SMP-24-015}

\providecommand{\cmsTable}[1]{\resizebox{\textwidth}{!}{#1}}
\ifthenelse{\boolean{cms@external}}{%
    \providecommand{\cmsleft}{upper\xspace}
    \providecommand{\cmsLeft}{Upper\xspace}
    \providecommand{\cmsright}{lower\xspace}
    \providecommand{\cmsRight}{Lower\xspace}
}{%
    \providecommand{\cmsleft}{left\xspace}
    \providecommand{\cmsLeft}{Left\xspace}
    \providecommand{\cmsright}{right\xspace}
    \providecommand{\cmsRight}{Right\xspace}
}
\newlength\cmsTabSkip\setlength{\cmsTabSkip}{2ex}

\newcommand{\ttZ}{\ensuremath{\ttbar\PZ}\xspace}
\newcommand{\tXZ}{\ensuremath{\PQt\PX\PZ}\xspace}
\newcommand{\ttV}{\ensuremath{\ttbar\PV}\xspace}
\newcommand{\ttH}{\ensuremath{\ttbar\PH}\xspace}
\newcommand{\tWZ}{\ensuremath{\PQt\PW\PZ}\xspace}
\newcommand{\tZq}{\ensuremath{\PQt\PZ\PQq}\xspace}
\newcommand{\WZ}{\ensuremath{\PW\PZ}\xspace}
\newcommand{\ZZ}{\ensuremath{\PZ\PZ}\xspace}

\newcommand{\abseta}{\ensuremath{\abs{\eta}}\xspace}
\newcommand{\DeepJet}{\ensuremath{\textsc{DeepJet}}\xspace}

\newcommand{\pp}{\ensuremath{\Pp\Pp}\xspace}
\newcommand{\WW}{\ensuremath{\PW\PW}\xspace}
\newcommand{\WWW}{\ensuremath{\PW\PW\PW}\xspace}
\newcommand{\WZZ}{\ensuremath{\PW\PZ\PZ}\xspace}
\newcommand{\WWZ}{\ensuremath{\PW\PW\PZ}\xspace}
\newcommand{\ZH}{\ensuremath{\PZ\PH}\xspace}
\newcommand{\WH}{\ensuremath{\PW\PH}\xspace}
\newcommand{\ZZZ}{\ensuremath{\PZ\PZ\PZ}\xspace}
\newcommand{\VVV}{\ensuremath{\PV\PV\PV}\xspace}
\newcommand{\VVZ}{\ensuremath{\PV\PV\PZ}\xspace}

\newcommand{\Run}[1]{\mbox{Run~#1}\xspace}

\newcommand{\mttwo}{\ensuremath{m_{\textrm{T}2}}\xspace}

\cmsNoteHeader{SMP-24-015}
\title{Measurement of \texorpdfstring{\WWZ}{WWZ} and \texorpdfstring{\ZH}{ZH} production cross sections at \texorpdfstring{$\sqrt{s}=13$}{sqrt(s) = 13} and \texorpdfstring{13.6\TeV}{13.6 TeV}}

\date{\today}

\abstract{A measurement is presented of the cross section in proton-proton collisions for the production of two \PW bosons and one \PZ boson. It is based on data recorded by the CMS experiment at the CERN LHC at center-of-mass energies $\sqrt{s}=13$ and 13.6\TeV, corresponding to an integrated luminosity of 200\fbinv. Events with four charged leptons (electrons or muons) in the final state are selected. Both nonresonant \WWZ production and \ZH production, with the Higgs boson decaying into two \PW bosons, are reported. For the first time, the two processes are measured separately in a simultaneous fit. Combining the two modes, signal strengths relative to the standard model (SM) predictions of $0.75^{+0.34}_{-0.29}$ and $1.74^{+0.71}_{-0.60}$ are measured for $\sqrt{s}=13$ and 13.6\TeV, respectively. The observed (expected) significance for the triboson signal is 3.8 (2.5) standard deviations for $\sqrt{s}=13.6\TeV$, thus providing the first evidence for triboson production at this center-of-mass energy. Combining the two modes and the two center-of-mass energies, the inclusive signal strength relative to the SM prediction is measured to be $1.03^{+0.31}_{-0.28}$, with an observed (expected) significance of 4.5 (5.0) standard deviations.}

\hypersetup{
pdfauthor={CMS Collaboration},%
pdftitle={Measurement of WWZ and ZH production cross sections at sqrt(s) = 13 and 13.6 TeV},%
pdfsubject={CMS},
pdfkeywords={CMS, electroweak, Higgs, Z boson, W boson}
}

\maketitle

The standard model (SM) of particle physics predicts interactions between massive gauge bosons (\PW and \PZ bosons) via triple and quartic gauge couplings. The existence of these couplings can result in the production of multiple massive gauge bosons in proton-proton (\pp) interactions.
Studies of these processes
at the CERN LHC test the SM predictions and probe for new physics,
whose contributions
could lead to enhancements or reductions of the multiboson production rates with respect to the expectations.

Using \pp data at $\sqrt{s}=13\TeV$,
corresponding to an integrated luminosity of 137\fbinv,
the
CMS Collaboration~\cite{CMS:2008xjf,CMS:2023gfb} has
previously observed the \VVV production process
(where \PV refers to a \PW or \PZ boson)
and also obtained evidence for \WWZ individually~\cite{CMS:2020hjs}; these final states include contributions from processes with an on-shell Higgs (\PH) boson.
Additionally, the ATLAS Collaboration~\cite{ATLAS:2008xda} has observed at $\sqrt{s}=13\TeV$ the \VVZ production process
 and reported evidence
for \WWZ using a data set of 140\fbinv~\cite{ATLAS:2024nab}.

\begin{figure}[ht!]
\centering
\includegraphics[width=0.3\textwidth]{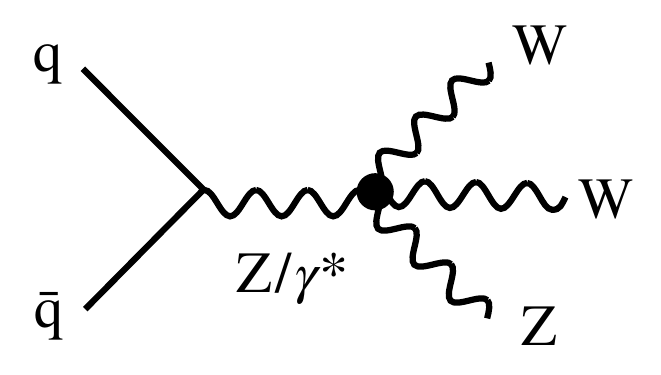}%
\hspace{0.1\textwidth}%
\includegraphics[width=0.3\textwidth]{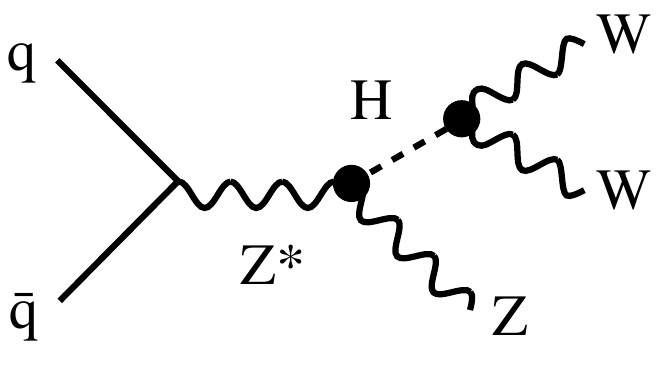}
\caption{%
    Example Feynman diagrams for nonresonant \WWZ (\cmsleft) and \ZH (with $\PH\to\WW$) production (\cmsright).
    The quartic gauge coupling is indicated in the \cmsleft diagram, and the gauge-Higgs couplings are indicated in the \cmsright diagram.
}
\label{fig:feynman_signal}
\end{figure}

The analysis described in this Letter uses data collected by the CMS Collaboration in
two separate running periods at slightly different center-of-mass energies.
The 2016--2018 period (corresponding to 138\fbinv at
$\sqrt{s}=13\TeV$) is referred to as ``\Run2'', and the 2022--2023 period
(corresponding to 62\fbinv at $\sqrt{s}=13.6\TeV$) is referred to as ``\Run3''.
With the two combined data sets, and concentrating on final states with four charged leptons (\Pe or \PGm),
we are now reporting the most sensitive measurement of
the \WWZ process to date, which also results in the first evidence for triboson production at 13.6\TeV.
We also report the first simultaneous study that separates nonresonant \WWZ production from \ZH ($\PH\to\WW$) production.  Representative Feynman diagrams for these two
processes
are displayed in Fig.~\ref{fig:feynman_signal}.

The $\pp\to\ZH$ process has been studied in a variety of final states by the CMS and ATLAS Collaborations.
The \Run2 measurements result in signal strengths
$\mu(\ZH) = 1.29^{+0.22}_{-0.25}$ by CMS~\cite{CMS:2022dwd}, and $0.96^{+0.22}_{-0.21}$ by ATLAS~\cite{ATLAS:2022vkf}.
Signal strengths are defined as the ratios of the measured cross sections to the central value of their SM prediction \cite{CMS:2020hjs,MadGraph5_aMCatNLO,LHCHiggsCrossSectionWorkingGroup:2016ypw}; the uncertainty in the cross section prediction is not included.

The CMS apparatus
is a multipurpose, nearly hermetic detector, designed to trigger on~\cite{CMS:2020cmk,CMS:2016ngn}
and identify electrons, muons, photons, and (charged and neutral)
hadrons~\cite{CMS:2020uim,CMS:2018rym,CMS:2014pgm}. A global ``particle-flow''
(PF) algorithm~\cite{CMS:2017yfk} aims to reconstruct all individual particles
in an event, combining information provided by the all-silicon inner tracker and by the
crystal electromagnetic (ECAL) and brass-scintillator
hadron calorimeters (HCAL),
operating inside a 3.8\unit{T} superconducting solenoid, with data from the gas-ionization
muon detectors embedded in the flux-return yoke outside the solenoid.
The reconstructed particles are used to build \PGt-leptons and jets,
as well as to measure the missing transverse momentum vector
(\ptvecmiss)~\cite{CMS:2018jrd,CMS:2016lmd,CMS:2019ctu}, defined as the
negative of the vector sum of the transverse momenta (\pt) of all particles.

The energy of photons is obtained from the ECAL measurement. The energy of electrons is determined from a combination of the electron momentum at the primary interaction vertex as determined by the tracker, the energy of the corresponding ECAL cluster, and the energy sum of all bremsstrahlung photons spatially compatible with originating from the electron track. The energy of muons is obtained from the curvature and direction of the corresponding track. The energy of charged hadrons is determined from a combination of their momentum measured in the tracker and the matching ECAL and HCAL energy deposits, corrected for the response function of the calorimeters to hadronic showers. Finally, the energy of neutral hadrons is obtained from the corresponding corrected ECAL and HCAL energies.

Jets are reconstructed from PF candidates, using the anti-\kt algorithm with a distance parameter of 0.4~\cite{Cacciari:2008gp,Cacciari:2011ma}. Jet
energies are corrected for instrumental effects and contributions from additional \pp interactions within the same or nearby bunch crossings (pileup)~\cite{Sirunyan:2020foa}.
The pileup per particle identification algorithm (PUPPI)~\cite{Sirunyan:2020foa,Bertolini:2014bba} can be used to mitigate the effect of pileup at the reconstructed particle level, combining local shape information, event pileup properties, and tracking information.
For the study described in this Letter, the PUPPI algorithm is used to improve the reconstruction of jets in the
\Run3 data set and
\ptvecmiss in both the \Run2 and \Run3 data sets.

The four-lepton data set used in this study is collected with dilepton triggers, resulting in a trigger efficiency of
100\% for events with all four leptons within the \pt
and pseudorapidity ($\eta$)
acceptance specified below.
Simulated Monte Carlo (MC) samples are used to estimate the SM predictions for the yields of signal processes. The \WWZ signal process is generated with the \MGvATNLO version 2.6 (\Run2) and 2.9 (\Run3)~\cite{MadGraph5_aMCatNLO,Frederix:2012ps,Alwall:2007fs} program at next-to-leading order (NLO) precision.
The \PW bosons for this sample are on-shell; the minimum invariant mass for the pair of leptons from the \PZ boson is 30\GeV. 
The quark- and gluon-initiated contributions to the \ZH signal process are modeled separately with the \POWHEG v2 generator at NLO in perturbative quantum chromodynamics (QCD)~\cite{POWHEG1,POWHEG2,POWHEG3,POWHEG_VV,Frixione:2007nw}.
The \ZH samples require the invariant mass for the pair of leptons from the \PZ boson to be greater than 10\GeV.
Background processes are also estimated with MC simulations, as well as with dedicated data control regions, as discussed below. Some of the background processes (\WW, \ZZ, \ttbar, single top, and $\PH\to\ZZ$)
are produced with \POWHEG to NLO accuracy in QCD, whereas others (\WZ, \WWW, \WZZ, \ZZZ, \ttV, \ttH, \WH, \tWZ, \tZq, and {\PW}+jets) are simulated to NLO accuracy in QCD using \MGvATNLO. The $\Pg\Pg\to\ZZ$ process is produced with the \MCFM v7.0.1 program~\cite{Campbell:1999ah}. The NNPDF3.1 next-to-NLO (NNLO)~\cite{NNPDF3} sets of parton distribution functions are used in the simulations. The parton showering and hadronization are modeled with \PYTHIA~\cite{Sjostrand:2014zea,Bierlich:2022pfr} (version 8.240 for the \Run2 samples, and version 8.306 for the \Run3 samples) with
the CP5 underlying event tune~\cite{Sirunyan:2019dfx}. Finally, the detector response is simulated with the \GEANTfour~\cite{GEANT4:2002zbu} package.
The cross sections for all \VVV processes are normalized to the NLO values obtained from the \MGvATNLO generator~\cite{MadGraph5_aMCatNLO,Frederix:2012ps,Alwall:2007fs}. For each contribution to the \ZH signal process, the cross sections recommended by the LHC Higgs Working Group~\cite{Karlberg:2024zxx} are used.

The main physics objects used in the analysis are leptons (\Pe and \PGm),
escaping particles (\ptvecmiss), and jets.
We target final states where the two \PW bosons and the \PZ boson decay to leptons,
either directly or via intermediate tau-lepton decays,
and thus we select events with exactly four isolated leptons in the final state.
The leading (subleading) lepton in the event is required to have $\pt>25$ (15)\GeV, and the
\pt of the third and fourth lepton is required to have $\pt>10\GeV$.
Muons and electrons are required to have $\abseta<2.4$ and 2.5, respectively.
We demand that two of the four leptons be consistent with originating from the decay of an on-shell \PZ
boson: These two leptons, referred to as ``\PZ candidate leptons'',
are required to be of the same flavor, to have opposite charges, and to form an invariant mass
within 10\GeV of the known \PZ boson mass~\cite{ParticleDataGroup:2024cfk}.
The remaining two leptons must have opposite charges, and are referred to as
``\PW candidate leptons''.
To minimize background processes that include top quarks, \eg,
$\pp\to\ttZ$, we reject
events with at least one jet of $\pt>20\GeV$ and $\abseta<2.4$ if the jet is
consistent with originating from bottom quarks, as determined by the
\textsc{DeepJet} discriminator~\cite{BTV-16-002, Bols:2020bkb, CMS-DP-2018-058}.

Events are categorized according to the flavors of the \PW candidate leptons as ``opposite flavor'' (OF: $\Pe\PGm$) or ``same flavor'' (SF: $\Pe\Pe$ or $\PGm\PGm$).
This separation is performed because of the different background compositions
in the SF and OF channels.
Selected events in the SF channel are predominantly \ZZ events that decay fully leptonically, with one of the two \PZ bosons being off-shell.
On the other hand, the OF channel selection includes sizable contributions from $\pp\to\ttZ$,
$\pp\to\tWZ$, and $\pp\to\ZZ$
where in this last case one of the two \PZ bosons decays to \PGt leptons
that subsequently decay to \Pe or \PGm.

In the SF channel, the \PW candidate leptons are required to form an invariant mass
more than 10\GeV away from the known mass of the \PZ boson;
additionally, the \mttwo variable~\cite{LESTER1999, Barr_2003}
(which is constructed from the
transverse momenta of the \PW candidate leptons and the \ptvecmiss)
must be greater than
25\GeV.  This last requirement is effective at reducing the \ZZ background, which tends to produce events with softer leptons, less \ptmiss, and smaller \mttwo values (compared to the non-resonant \WWZ signal).
Contributions from events with low-mass resonances are suppressed by
rejecting events where a pair of opposite-charge
leptons forms an invariant mass below 12\GeV.
The event selection efficiency including branching fractions into electrons and muons is about 14 and 12\% for nonresonant \WWZ and \ZH, respectively, with some dependence on the run period due to small variations in the run conditions and reconstruction algorithms.

To better distinguish signal processes from background processes, and to distinguish the two signal processes
from each other, a boosted decision tree (BDT) multiclassifier~\cite{hoecker2009tmvatoolkitmultivariate} is
trained on signal and background MC events.
The multiclassifier is based on 27 event kinematic properties, such as invariant masses of combinations of leptons, \ptmiss, lepton \pt, angular separations between objects (leptons and combinations of leptons, \ptmiss, jets).
The multiclassifier assigns three scores to each event: a \WWZ score, a \ZH score, and a background score.  The three scores
are transformed using the \texttt{softmax} function
in such a way that the sum of all scores for a given
event is equal to one.  This then allows the individual scores to be interpreted as probabilities
of the event to originate from one of the three sources.
The most important variables in separating signals from backgrounds and the two signals from each other
are the invariant masses of the four leptons and of the \PW leptons, the \ptmiss, the
azimuthal angle between the four-lepton system and \ptvecmiss, and the $\eta$-$\phi$ separation between the two \PW leptons.

Signal regions (SRs)
are then constructed by binning in the three BDT scores.
For \Run2, we define eight SR bins per channel (OF or SF), and for
\Run3, which has about one half the integrated luminosity of \Run2,
the number of bins per channel is reduced to four.

We use dedicated control regions (CRs) to aid in the estimate of the leading background contributions,
which consist of \ZZ, \ttZ, and \tWZ production.
For the \ZZ
process, we define one CR per run period, with requirements identical to the SF signal selection except that the invariant mass of both pairs of leptons are required to be within the \PZ boson mass window and the \mttwo requirement is not applied.
We construct CRs for \ttZ and \tWZ (referred to as the \tXZ CR) by requiring the presence of at least one \PQb-tagged jet, defining a total of eight CR bins by splitting events by SF/OF channels, run period, and \PQb-tagged jet multiplicity (one or more than one).
In addition, for the SF \tXZ CRs, we replace the $\mttwo>25\GeV$ requirement with a $\ptmiss>80\GeV$
requirement, since the latter
is more effective at reducing the contributions from the \ZZ process in these dedicated CRs.

\begin{figure*}[!htp]
\centering
\includegraphics[width=\textwidth]{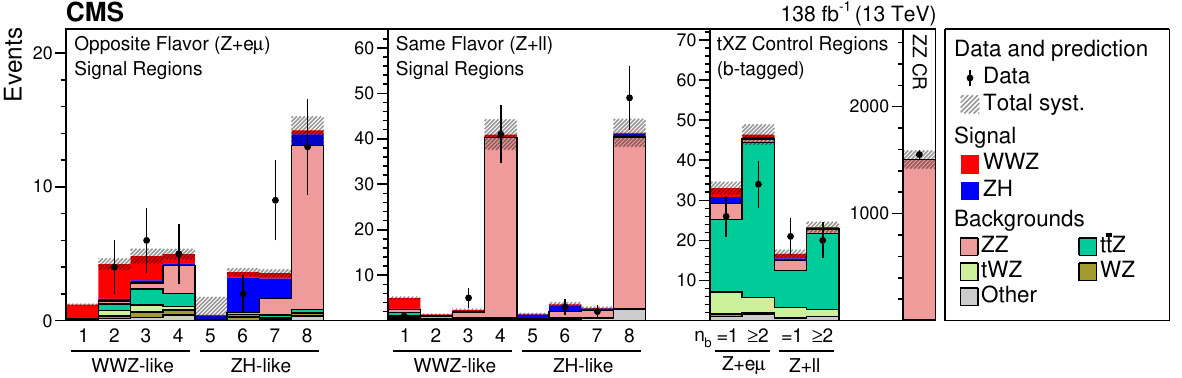} \\[1ex]
\includegraphics[width=\textwidth]{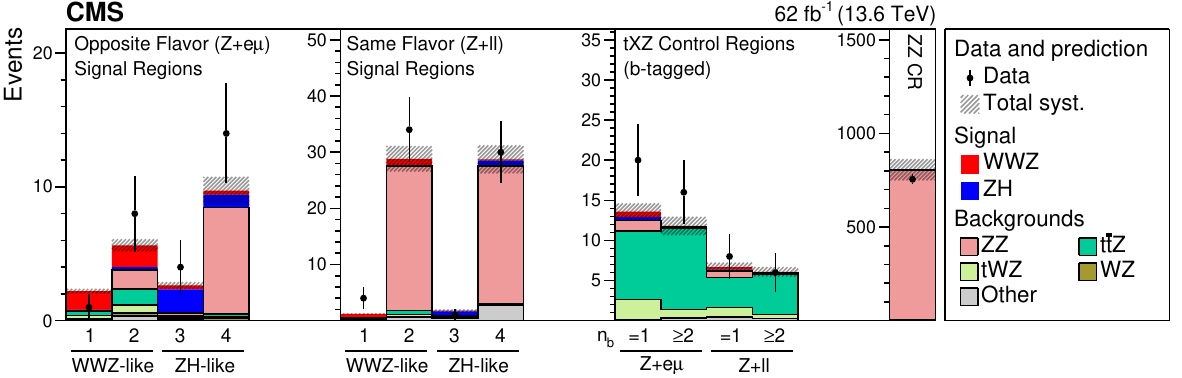}
\caption{%
    Comparison of the numbers of observed to SM-predicted events for each of the bins included
    in the fit for \Run2 (upper row) and \Run3 (lower row).  The SM expectations are shown before performing the fit.
    The horizontal axis legend also indicates ``\WWZ-like'' and ``\ZH-like'' bins defined based on whether the \WWZ BDT
    score is higher or lower than the \ZH score. The error bars on the data points represent the statistical uncertainties. The shaded bands represent the total contributions of the systematic uncertainties.
}
\label{fig:results:run2:bdt_bins}
\end{figure*}

Smaller background contributions include triboson processes (\WWW, \WZZ, \ZZZ), as well as events
with at least one lepton candidate that is not from \PW or \PZ boson decay.
These backgrounds originate from diboson production
(\WZ, same- and opposite-sign \WW),
top quark+\PX, Higgs boson,
\ttbar, single top, and {\PV}+jets processes, and
are estimated with MC simulation with appropriate cross section uncertainties applied
in the fitting procedure described below.
The event yields and the SM prediction for signal and background in the CRs and SRs are displayed in
Fig.~\ref{fig:results:run2:bdt_bins}, separately for the two run periods.

The yields for signal (\WWZ and \ZH) and background are extracted from a simultaneous extended
maximum likelihood fit to the bins of Fig.~\ref{fig:results:run2:bdt_bins}.
The \ZZ normalization and the {\ttZ}+\tWZ normalization are allowed to float freely in the fit, since they are quite effectively constrained by the control
regions.  The normalizations of the other backgrounds are taken from standard model MC expectations,
with appropriate systematic uncertainties.
In the simultaneous fit, uncertainties and correlations are
handled automatically by the statistical analysis tool, \textsc{combine}~\cite{CMS:2024onh},
with a consistent handling of signal contributions to the control regions.

We account for uncertainties in the fit arising from experimental sources,
as well as those associated with the theoretical modeling, by including nuisance parameters in the likelihood.
Amongst these uncertainties, the most important ones are those associated
with the extraction of the normalization of the main background processes from the control regions
and the efficiency of the lepton identification requirements (about 2\% per electron and 1\% per muon).
Other experimental sources of uncertainties include the uncertainty associated with the CMS integrated
luminosity measurements for \Run2 (1.6\%)~\cite{CMS:2021xjt,CMS-PAS-LUM-17-004,CMS-PAS-LUM-18-002} and \Run3 (1.4\%)~\cite{CMS:LUM-22-001}, the simulation of pileup (2\%),
the uncertainty in the jet energy scale and resolution and its impact on \ptmiss (2\%),
and the uncertainty in the \PQb tagging efficiency (1\%).
For theoretical modeling, we include uncertainties in the acceptances of the various processes
associated with the MC modeling of parton
showers and initial and final state radiation, as well as the assumptions on the renormalization and factorization
scales used in the MC generation of the various samples (typically 2\%).  Finally, we assign
a 30\%
uncertainty to the normalization of the small subleading background processes
whose normalizations are not extracted from the fit to the CRs.

\begin{figure}[htp!]
\centering
\includegraphics[width=0.48\textwidth]{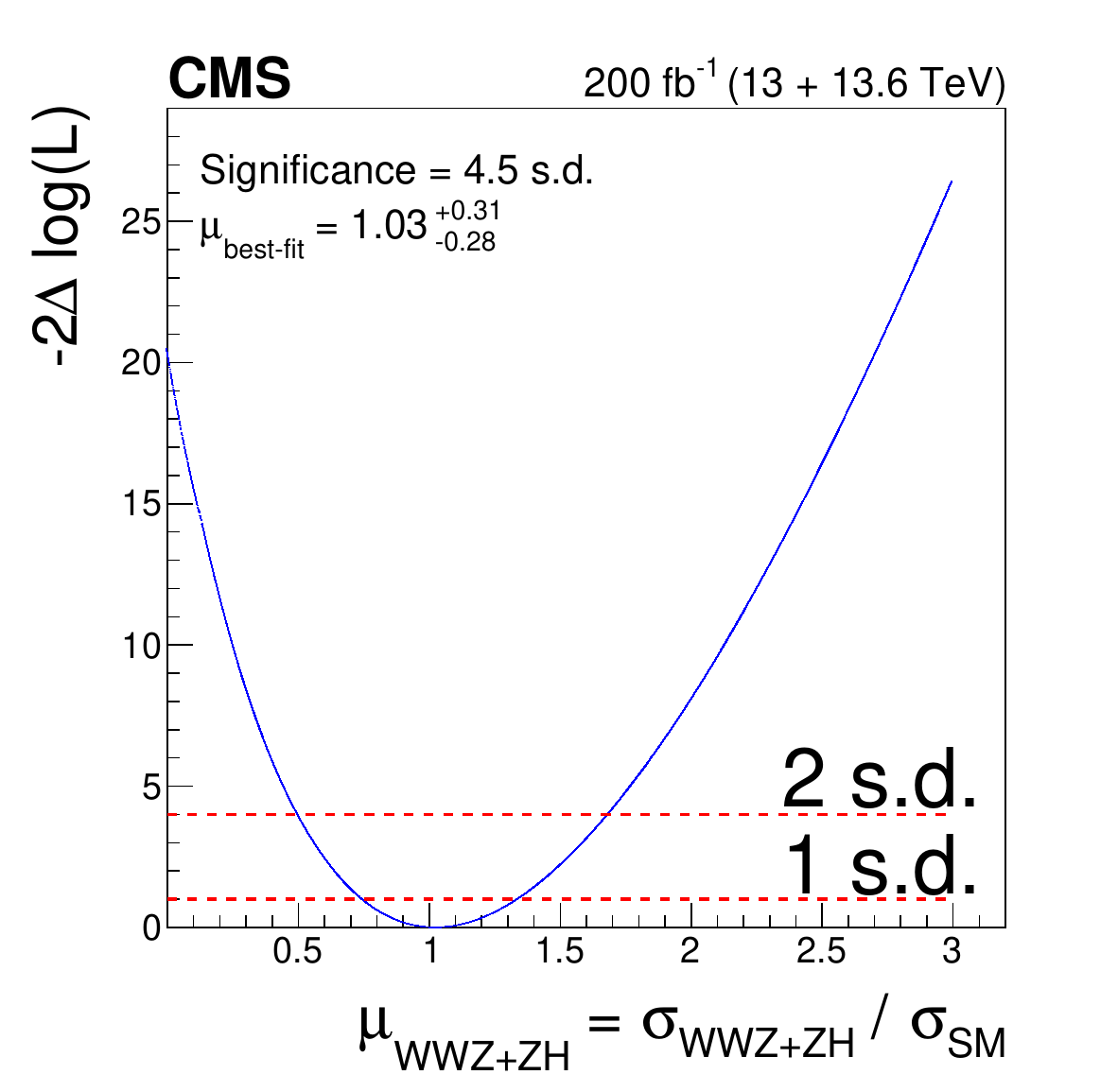}%
\hfill%
\includegraphics[width=0.48\textwidth]{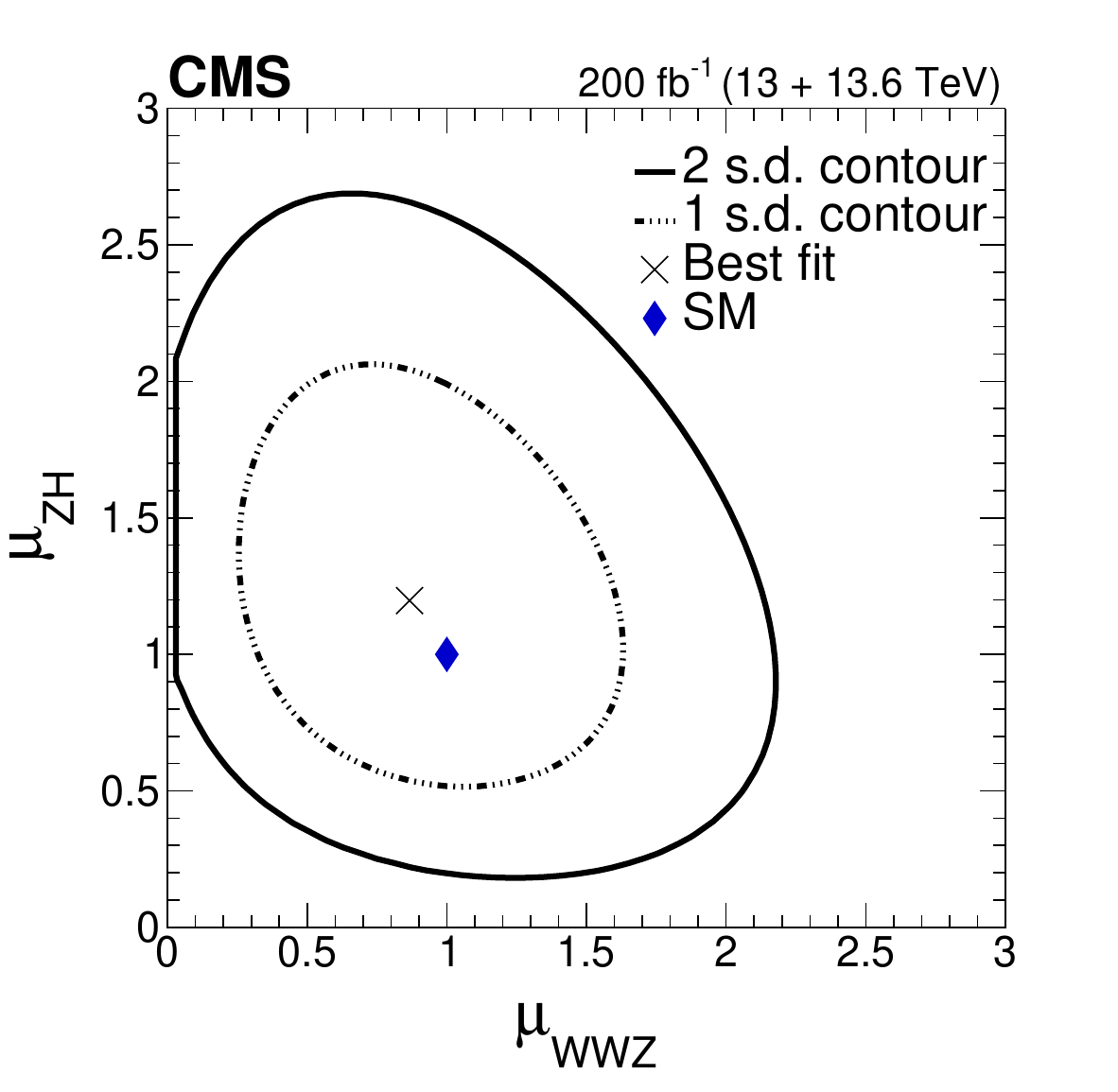}
\caption{%
    Likelihood scans for the combined
    \Run2 and \Run3 data sets.  \cmsLeft: One-dimensional likelihood scan as a function of the inclusive \WWZ and \ZH signal strength parameter ($\mu_{\WWZ{+}\ZH}$).
    \cmsRight:
    Two-dimensional scan as a function of the individual \WWZ and \ZH signal strength parameters. For the two-dimensional scan, the one and two standard deviation (s.d.) contours
    are indicated by the dotted and solid lines, respectively.  The correlation coefficient is $-0.23$.
    In the signal strengths reported here, the uncertainty of the predicted cross section is not included.
}
\label{fig:results:comb:1D2Dscan}
\end{figure}

The fitted yields are used to compute the signal strength and statistical significance of the
two signal processes, \WWZ and \ZH, in both data-taking periods, separately by process and period,
inclusive across processes, and inclusive across periods. Signal strengths and cross sections are shown in Table~\ref{tab:results:unblinded:xsecs}; for the \ZH signal process, we report the cross section
times the $\PH\to\PW\PW$ branching fraction~\cite{LHCHiggsCrossSectionWorkingGroup:2013rie}.
A one-dimensional profile likelihood scan of the inclusive total (\WWZ and \ZH) signal strength and a two-dimensional scan of the \WWZ and \ZH signal strengths are shown in Fig.~\ref{fig:results:comb:1D2Dscan}.
From the one-dimensional scan, we measure the inclusive total signal strength to be $1.03^{+0.31}_{-0.28}$, with an observed (expected) significance of 4.5 (5.0)
standard deviations.

\begin{table*}[htp!]
\centering
\topcaption{%
    Measured signal strengths and cross sections for the \WWZ, \ZH (with $\PH\to\WW$) and inclusive ($\WWZ{+}\ZH$) processes at $\sqrt{s}=13$ and 13.6\TeV. For the \ZH process, the cross section times the $\PH\to\WW$ branching fraction~\cite{LHCHiggsCrossSectionWorkingGroup:2013rie} is reported. The last column shows the SM expectation; the \WWZ cross sections are calculated to NLO precision while the \ZH cross sections are calculated at NNLO. In the signal strengths reported here, the uncertainty of the predicted cross section is not included.
}
\label{tab:results:unblinded:xsecs}
\renewcommand{\arraystretch}{1.5}
\cmsTable{\begin{scotch}{lcccccc}
    Era & Process & $\mu = \sigma_{\text{obs}}/\sigma_{\text{SM}}$ & $\sigma_{\text{obs}}$ (fb) & $\sigma_{\text{SM}}$ (fb) \\
    \hline
    \multirow{3}{*}{\Run2 ($\sqrt{s}=13\TeV$)}
    & \WWZ only & $0.52^{+0.50}_{-0.41}\stat^{+0.13}_{-0.12}\syst$ & $97^{+91}_{-75}\stat^{+24}_{-21}\syst$ & 184 \cite{MadGraph5_aMCatNLO} \\
    & \ZH only & $1.01^{+0.62}_{-0.51}\stat^{+0.14}_{-0.08}\syst$ & $190^{+120}_{-100}\stat^{+30}_{-20}\syst$ & 189 \cite{LHCHiggsCrossSectionWorkingGroup:2016ypw} \\
    & Inclusive & $0.75^{+0.32}_{-0.28}\stat^{+0.10}_{-0.08}\syst$ & $280^{+120}_{-110}\stat^{+40}_{-30}\syst$ & 373 \\[\cmsTabSkip]
    \multirow{3}{*}{\Run3 ($\sqrt{s}=13.6\TeV$)}
    & \WWZ only & $1.91^{+1.12}_{-0.94}\stat^{+0.25}_{-0.16}\syst$ & $380^{+220}_{-190}\stat^{+50}_{-30}\syst$ & 200 \cite{MadGraph5_aMCatNLO} \\
    & \ZH only & $1.57^{+1.12}_{-0.89}\stat^{+0.29}_{-0.14}\syst$ & $320^{+230}_{-180}\stat^{+60}_{-30}\syst$ & 202 \cite{Karlberg:2024zxx} \\
    & Inclusive & $1.74^{+0.67}_{-0.58}\stat^{+0.22}_{-0.15}\syst$ & $700^{+270}_{-230}\stat^{+90}_{-60}\syst$ & 402 \\
\end{scotch}}
\end{table*}

Since the BDT is very effective in separating \ZH from nonresonant \WWZ, fixing the \ZH cross section to its measured value~\cite{CMS:2022dwd} does not appreciably change the fitted value of the \WWZ signal strength and its uncertainty (less than 1\% for the signal strength and about 2\% for the uncertainty).

\begin{figure*}[!htp]
\centering
\includegraphics[width=0.8\textwidth]{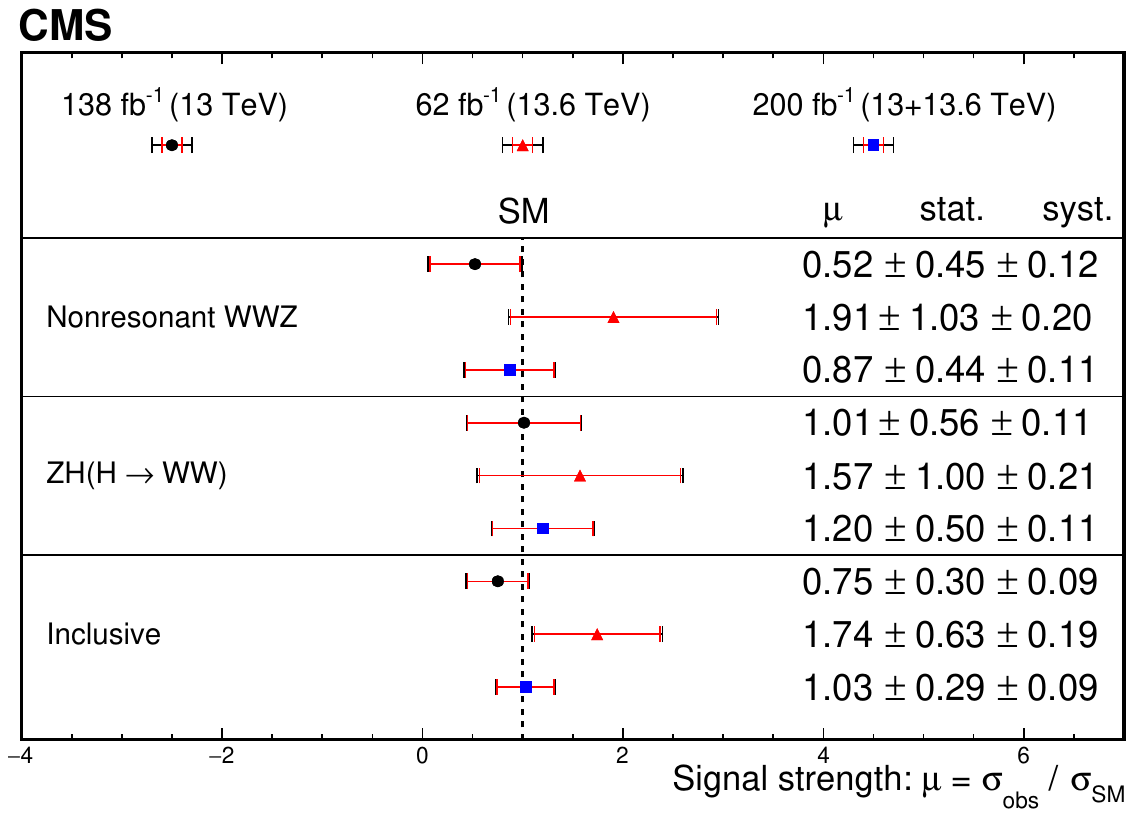}
\caption{%
    Measured values of the \WWZ, \ZH, and inclusive signal strength parameters.  Results are shown
    separately by run period as well as after combining the two data sets. The red error bars on the signal strengths represent the statistical uncertainty, and the black error bars represent the total uncertainty. In the signal strengths reported here, the uncertainty of the cross section prediction is not included.
}
\label{fig:results:comb:lep}
\end{figure*}

The results are summarized in Tables~\ref{tab:results:unblinded:xsecs} and \ref{tab:results_sig}.
In Fig.~\ref{fig:results:comb:lep}, the measured signal strengths are displayed by process and run period.
We note that with the size of the currently available data set, the statistical uncertainties are much larger than the systematic ones;
comparing the statistical-only uncertainty to the total uncertainty (\ie, statistical and systematical in quadrature) in each of the signal strength measurements, the inclusion of the systematic uncertainty increases the total uncertainty by less than 5\%.
Numerical values for figures are provided in the HEPData record for this analysis~\cite{hepdata}.

\begin{table}[!htp]
\centering
\topcaption{%
    Observed (expected) significance of the signal strength in units of standard deviations. The \WWZ-only and \ZH-only values are obtained by profiling the other signal process, together with all other nuisances.
}
\label{tab:results_sig}
\renewcommand{\arraystretch}{1.1}
\begin{scotch}{lccc}
    Era & Total & \WWZ & \ZH\\
    \hline
    \Run2 & 2.9 (4.4) & 1.3 (3.1) & 2.0 (2.6) \\
    \Run3 & 3.8 (2.5) & 2.5 (1.3) & 2.5 (1.7) \\
    Total & 4.5 (5.0) & 2.4 (3.3) & 3.1 (3.1) \\
\end{scotch}
\end{table}

In summary, in this Letter we have presented a measurement of the
\WWZ production cross section in proton-proton collisions in the four charged lepton (\Pe, \PGm) final state with a data set
corresponding to an integrated luminosity of 200\fbinv
collected by the CMS Collaboration at center-of-mass energies of 13 and 13.6\TeV.
The cross section relative to the standard model prediction is measured to be $1.03^{+0.31}_{-0.28}$,
representing the most precise measurement of this process to date.
The result also  provides
the first evidence for a triboson production process at 13.6\TeV.
Additionally, the nonresonant and \ZH (with the Higgs boson \PH decaying to \PW bosons) contributions are studied simultaneously for the first time.  Within the current uncertainties, our results are in agreement with standard model predictions.

\begin{acknowledgments}
We congratulate our colleagues in the CERN accelerator departments for the excellent performance of the LHC and thank the technical and administrative staffs at CERN and at other CMS institutes for their contributions to the success of the CMS effort. In addition, we gratefully acknowledge the computing centers and personnel of the Worldwide LHC Computing Grid and other centers for delivering so effectively the computing infrastructure essential to our analyses. Finally, we acknowledge the enduring support for the construction and operation of the LHC, the CMS detector, and the supporting computing infrastructure provided by the following funding agencies: SC (Armenia), BMBWF and FWF (Austria); FNRS and FWO (Belgium); CNPq, CAPES, FAPERJ, FAPERGS, and FAPESP (Brazil); MES and BNSF (Bulgaria); CERN; CAS, MoST, and NSFC (China); MINCIENCIAS (Colombia); MSES and CSF (Croatia); RIF (Cyprus); SENESCYT (Ecuador); ERC PRG, RVTT3 and MoER TK202 (Estonia); Academy of Finland, MEC, and HIP (Finland); CEA and CNRS/IN2P3 (France); SRNSF (Georgia); BMBF, DFG, and HGF (Germany); GSRI (Greece); NKFIH (Hungary); DAE and DST (India); IPM (Iran); SFI (Ireland); INFN (Italy); MSIP and NRF (Republic of Korea); MES (Latvia); LMTLT (Lithuania); MOE and UM (Malaysia); BUAP, CINVESTAV, CONACYT, LNS, SEP, and UASLP-FAI (Mexico); MOS (Montenegro); MBIE (New Zealand); PAEC (Pakistan); MES and NSC (Poland); FCT (Portugal);  MESTD (Serbia); MICIU/AEI and PCTI (Spain); MOSTR (Sri Lanka); Swiss Funding Agencies (Switzerland); MST (Taipei); MHESI and NSTDA (Thailand); TUBITAK and TENMAK (Turkey); NASU (Ukraine); STFC (United Kingdom); DOE and NSF (USA).
\end{acknowledgments}

\bibliography{auto_generated}
\cleardoublepage \appendix\section{The CMS Collaboration \label{app:collab}}\begin{sloppypar}\hyphenpenalty=5000\widowpenalty=500\clubpenalty=5000
\cmsinstitute{Yerevan Physics Institute, Yerevan, Armenia}
{\tolerance=6000
A.~Hayrapetyan, V.~Makarenko\cmsorcid{0000-0002-8406-8605}, A.~Tumasyan\cmsAuthorMark{1}\cmsorcid{0009-0000-0684-6742}
\par}
\cmsinstitute{Institut f\"{u}r Hochenergiephysik, Vienna, Austria}
{\tolerance=6000
W.~Adam\cmsorcid{0000-0001-9099-4341}, J.W.~Andrejkovic, L.~Benato\cmsorcid{0000-0001-5135-7489}, T.~Bergauer\cmsorcid{0000-0002-5786-0293}, K.~Damanakis\cmsorcid{0000-0001-5389-2872}, M.~Dragicevic\cmsorcid{0000-0003-1967-6783}, C.~Giordano, P.S.~Hussain\cmsorcid{0000-0002-4825-5278}, M.~Jeitler\cmsAuthorMark{2}\cmsorcid{0000-0002-5141-9560}, N.~Krammer\cmsorcid{0000-0002-0548-0985}, A.~Li\cmsorcid{0000-0002-4547-116X}, D.~Liko\cmsorcid{0000-0002-3380-473X}, I.~Mikulec\cmsorcid{0000-0003-0385-2746}, J.~Schieck\cmsAuthorMark{2}\cmsorcid{0000-0002-1058-8093}, R.~Sch\"{o}fbeck\cmsAuthorMark{2}\cmsorcid{0000-0002-2332-8784}, D.~Schwarz\cmsorcid{0000-0002-3821-7331}, M.~Shooshtari, M.~Sonawane\cmsorcid{0000-0003-0510-7010}, W.~Waltenberger\cmsorcid{0000-0002-6215-7228}, C.-E.~Wulz\cmsAuthorMark{2}\cmsorcid{0000-0001-9226-5812}
% ...
\par}
\vskip\cmsinstskip
\dag:~Deceased\\
$^{1}$Also at Yerevan State University, Yerevan, Armenia\\
$^{2}$Also at TU Wien, Vienna, Austria\\
% ...
\end{sloppypar}
\end{document}